\documentclass[12pt,preprint]{aastex}

\title{Wide-Angle Wind Driven Bipolar Outflows:  High Resolution Models
with Application to Source I of the Becklin-Neugebauer / Kleinmann-Low
OMC-I Region}

\author{Andrew Cunningham\altaffilmark{1}, Adam Frank\altaffilmark{1}} 
\affil{Bausch \& Lomb Hall, University of Rochester, Rochester, NY  14627}
\and
\author{Lee Hartmann\altaffilmark{2}} 
\affil{Harvard-Smithsonian Center for Astrophysics, 60 Garden Street, Cambridge, MA 02138}

\altaffiltext{1}{Department ofPhysics and Astronomy, University of Rochester}

\altaffiltext{2}{Harvard-Smithsonian Center for Astrophysics}

\shorttitle{High Mass YSO Molecular Outflow}

\shortauthors{Cunningham,Frank,Hartmann}
\received{2005 April 12}
\begin{document}
\begin{abstract}

We carry out high resolution simulations of the inner regions of a
wide angle wind driven bipolar outflow using an Adaptive Mesh
Refinement code. Our code follows H-He gas with molecular, atomic and
ionic components and the associated time dependent molecular chemistry
and ionization dynamics with radiative cooling. Our simulations
explore the nature of the outflow when a spherical wind expands into a
rotating, collapsing envelope.  We compare with key observational
properties of the outflow system of Source I in the BN/KL region.

Our calculations show that the wind evacuates a bipolar outflow cavity
in the infalling envelope.  We find the head of the outflow to be
unstable and that it rapidly fragments into clumps.  We resolve the
dynamics of the strong shear layer which defines the side walls of the
cavity.  We conjecture that this layer is the likely site of maser
emission and examine its morphology and rotational properties. The
shell of swept up ambient gas that delineates the cavity edge retains
its angular momentum. This rotation is roughly consistent with that
observed in the Source I $SiO$ maser spots.  The observed proper
motions and line-of-sight velocity are approximately reproduced by the
model.  The cavity shell at the base of the flow assumes an X-shaped
morphology which is also consistent with Source I.  We conclude that
the wide opening angle of the outflow is evidence that a wide-angle
wind drives the Source I outflow and not a collimated jet.

\end{abstract}


\section{Introduction}

Bipolar outflows are recognized as a fundamental component of the star
formation process for low and intermediate mass stars. The precise nature of
these outflows has received considerable attention and their
observational properties have been well characterized
(\cite{2000prpl.conf..867R}).  The nature of the mechanisms driving
the outflows has not, however, been determined.  While there is now a
consensus that the outflow is made up of swept-up ambient material
driven by a ``wind'' from the central source (a proto-star or young
star) there remains a debate over the form such a wind will take.  In
some models the star produces a well collimated jet on small scales
($<10 ~AU$) due (most likely) to the action of an accretion disk
(\cite{2000prpl.conf..759K}). Other models assume that the
proto-stellar wind is not as strongly collimated taking the form of a
``wide-angle wind'' (WAW). In many models the WAW will have a momentum
distribution such that wind has dense core in the polar direction
surrounded by less dense flow at lower latitudes
(\cite{2000prpl.conf..789S}). Thus WAWs are often envisioned as
having a jet-like component. The distinction between these models
plays an important role in the debate about star formation because the
nature of the WAW/jet is tied to the nature of the wind launching
mechanism at the base. Understanding how winds are driven from an
accretion disk (most likely via a coupling of magnetic fields and
rotation) is a problem of great importance both because jets/winds are
ubiquitous in astrophysical environments and because these flows may
carry away a significant fraction of the disk's angular momentum
(\cite{2000prpl.conf..759K}).  Thus the distinction between tightly
collimated jets and wide-angle winds informs the debate about the
nature of jet launching and accretion disk physics. Since it is
difficult to directly observe the wind launching regions (see
\cite{2005A&A...432..149W} ), the properties of large scale bipolar
outflows can aid in understanding the nature of their initial
conditions.

In addition to questions concerning the nature of the driving winds,
bipolar outflows may play an important role in modifying their natal
environments. The energy and momentum budgets associated with the
aggregate of outflows in a young cluster can be large enough to either
power turbulence in the cloud from which the cluster was born or, in
some cases, unbind some fraction of the cloud material
(\cite{2003RMxAC..15..123A}).  Observational and theoretical studies
have yet to address this issue in its full complexity and so the role
of outflows as environmental factors determining cloud properties
remains unclear. Recent studies of the NGC 1333 region have shown that
fossil bipolar outflows or ``cavities'' which remain after the central
source has turned off provide a significant coupling agent linking
wind momenta to the cloud (\cite{Quillenea2004}).  This work
emphasizes the importance of understanding outflow properties in terms
of momentum transfer processes. Understanding how the these properties
operate over the entire outflow history will be critical to addressing
their impact on the cloud both on intermediate scale ($\sim 10^{16} -
10^{17}~cm$) and large scales ($\ga .1~pc$).

We note that while considerable progress has been made in
understanding star formation for isolated low mass stars, the
formation of high-mass stars remains less clear (see the excellent
review by \cite{2003ASPC..287..333S}). There is increasing
observational work characterizing outflows in high mass star forming
environments however fundamental questions such as the role of
accretion disks, magnetic fields remain to be definitively
answered. Thus there remains considerable work to be done in the study
of bipolar outflows in the context of more massive stars.  It is
noteworthy that the distinction between jet driven and WAW driven
outflows is even less clear in high mass stars since radiation
pressure is capable of producing a significant stellar wind in many
cases.

In this context observations of an organized distribution of $Si0$ and
$H_2O$ maser emitting spots near Source I in the Orion BN/KL nebula
are of particular interest. The exact nature of Source I is unclear,
because it is so heavily embedded
\citep{greenhill,2002ASPC..267..357C}.  $SiO$ masers are distributed
along an {\bf X}-shaped locus of clumps extending 20 to 70 AU from the
continuum radio source \citep{snyder, wright83, lane,
plambeck}. Subsequent observations by
\cite{menten,greenhill98,doeleman} have established that the maser
sources appear to part of an outflow from source I. The maser proper
motions lie in the range of 10 to 23 $km~s^{-1}$, oriented primarily
along the limbs of the {\bf X}, with systematically red-shifted and
blue-shifted lobes about a southeast-northwest symmetry axis.

A number of models have been proposed to explain the pattern of maser
proper motion and line of sight velocity (\cite{plambeck, greenhill98}).
Recently \cite{greenhill, greenhill05} have presented a model based on
high resolution radio observations where the outflow is
northeast-southwest so that the masers are situated along a biconical
outflow with a wide opening angle.  \cite{greenhill} note a bridge of
maser emission connecting the southern and western arms of the bicone,
with a clear velocity gradient that is consistent with the edge of a
rotating disk. Orienting the outflow northeast to southwest makes the
ejection perpendicular to the disk, as seen in many low-mass young
stellar objects, and suggests that the $H_2O$ masers seen on larger
scales are also oriented with, and thus probably produced by, the same
outflow as the SiO masers \citep{greenhill98}.

The radial velocity difference between the arms suggests rotation.  By
requiring that this rotational motion be consistent with gravitational
binding, \cite{greenhill} estimated the dynamical mass of source I to
be $\sim 6~M_{\sun}$.

What is the origin of this rotational motion?  One possibility is that
there are streams of slow-moving molecular ejecta from the outer edges
of the disk (L. Greenhill, personal communication).  The other
possibility is that this rotation reflects the angular momentum in the
ambient medium. The presence of rotation and outflow in the Source I
system makes it an interesting test case for models of proto-stellar
winds interacting with the surrounding media on relatively small
scales ($<70~AU$).

Motivated by these observations of source I, in this paper we present
a numerical study of the interaction between a fast irrotational wind
from a central source with an infalling, rotating protostellar
envelope.  This work is a continuation of an going study of outflow
properties formed via wind/infalling envelope interactions.  In
previous works we have explored the roles of inflow ram pressure on
outflow collimation (\cite{delamarter}) as well as the role of
toroidal magnetic fields in shaping the outflow (\cite{gardiner}). In
this work we focus on the walls of the outflow where the development
of a shear layer between the infalling ambient material and the
outflowing wind material is the site of strong mixing and could be the
site of maser emission.  We explore the dynamics of the shell walls
and show that the velocities in this region are reasonably consistent
with the observations of Source I.

Our use of an adaptive mesh refinement allows us to achieve high
levels of resolution within the walls of the bipolar outflow.  By
capturing the hydrodynamics of these strongly cooling flows we can
address some open issues in the physics of molecular outflows.  In
particular we have been able to observe instabilities occurring at the
top, or head of the outflow, as well as partially resolve the flow
pattern which occurs as shocked infalling material flows past shocked
outflowing wind.  The latter question is some importance as there has
been debate in previous works on the subject as to the nature of the
mixing in this region and the direction of the final bulk momentum
\citep{delamarter,lee,shu,wilkin}. Thus our paper addresses generic
issues related to strongly cooling wind blown bubbles as well as some
specific issues related directly to nature of the outflow from Source
I.

In section II we discuss the numerical model and initial conditions
used for the simulations as well as the assumptions and
simplifications which support the model.  In appendix A more detail of
our micro-physics is provided. In section III we present our
results. In section IV we discuss the results in light of observations
of Source I and general considerations of molecular outflows.  In
section V we present our conclusions.

\section{Method and Model} \subsection{Numerical Code} We have carried out
a series of radiative hydrodynamic simulations of an isotropic wind
interacting with a rotating, infalling envelope. The micro-physics
of H and He ionization, $H_2$ chemistry and optically thin cooling
have been included. Our simulations are carried out in 2.5D
(i.e. cylindrical symmetry) using the AstroBEAR adaptive mesh
refinement (AMR) code. AMR allows high resolution to be achieved only
in those regions which require it due to the presence of steep
gradients in critical quantities such as gas density. The hydrodynamic
version of AstroBEAR has been well tested on variety of problem in 1,
2 and 2.5D \citep{pol,var} The system of equations integrated are:
\[d_t Q + \bigtriangledown \cdot F = S_{geom} + S_m + S_{grav}\] where
the vector of conserved quantities $Q$, the flux function $F$, the
geometric source terms $S_{geom}$, the micro-physical source terms
$S_m$, and the central gravitational source terms $S_{grav}$ are given
as:
\[
Q = \left[ \begin{array}{c}
\rho \\ \rho v_r \\\rho v_z \\ \rho v_{\theta} \\ E \\
\rho_{H_2} \\ \rho_{HI} \\ \rho_{HII} \\ \rho_{HeI} \\ \rho_{HeII}
\end{array}\right],
F_* = v_* \left[ \begin{array}{c}
\rho \\ \rho v_r \\\rho v_z \\ \rho v_{\theta} \\ E+P \\
\rho_{H_2} \\ \rho_{HI} \\ \rho_{HII} \\ \rho_{HeI} \\ \rho_{HeII}
\end{array}\right],
S_{geom}= \left[ \begin{array}{c}
\rho v_r \\ \rho v_r^2-\rho v_{\theta}^2 \\ \rho v_r v_z \\ 2 \rho v_r v_{\theta} \\ v_r(E+P) \\
\rho_{H_2} v_r \\ \rho_{HI} v_r \\ \rho_{HII} v_r \\ \rho_{HeI}
v_r \\ \rho_{HeII} v_r
\end{array}\right],
\]
\[
S_{m}= \left[ \begin{array}{c}
0 \\ 0 \\ 0 \\ 0 \\ -\Lambda \\
\mu_{H_2}(R_{H_2}-D_{H_2}) \\ \mu_{HI}(2(D_{H_2}-R_{H_2}) +
D_{HII}-R_{HII}) \\ \mu_{HII}(R_{HII}-D_{HII}) \\
\mu_{HeI}(D_{HeII}-R_{HeII}) \\ \mu_{HeI}(R_{HeII}-D_{HeII})
\end{array}\right],
S_{grav}= \frac{-GM}{(r^2+z^2)^{3/2}} \left[ \begin{array}{c}
0 \\ \rho r \\ \rho z \\ 0 \\ \rho v_r r + \rho v_z z \\
0 \\ 0 \\ 0 \\ 0 \\ 0
\end{array}\right],
\]
where $\rho$ is the gas density, $v_r$, $v_z$ and $v_\theta$ are the
components of the velocity, $E$ is the total energy, $P$ is the gas
pressure and $\mu_*$ is the molecular weight of each species.  The
code tracks $H_2$, $HI$, $HII$, $HeI$, and $HeII$ densities separately
using the self-consistent multifluid advection method of
\cite{plewa}. The $R_*$ and $D_*$ terms indicate recombination
dissociation rates respectively. The cooling, dissociation and
recombination rates included in the microphysical source term, $S_m$,
are given in appendix A. The local value of the adiabatic index,
$\gamma$ and the mean molecular weight of the gas is dependent on the
local gas composition.  In our code these values are taken as
piecewise constant across grid cells.  We have neglected the effects
of $H_2$ recombination heating and doubly ionized He.  The effect of
these processes are small for the conditions considered here.

The code employs an exact hydrodynamic Riemann solver. A spatial and
temporal second order accurate wave propagation scheme
(\cite{leveque}) is used to advance the solution of the source-free
Euler equations. Our code also maintains spatial and temporal second
order accuracy in pressure using the method of \cite{balsara}. The
geometric and micro-physical source terms are handled separately from
the hydrodynamic integration using an operator split approach.  The
source term $S=S_{geom}+S_m+S_{grav}$ is integrated using an implicit
fourth-order Rosenbrock integration scheme for stiff ODE's. We note
that adaptive mesh refinement has been particularly useful in
resolving the neighborhood of thin shock bounded cavity walls that are
prevalent in wind blown bubble environments.

\subsection{Model Parameters and Assumptions} High mass YSO outflows
typically exhibit wider opening angles than their low mass
counterparts \citep{konigl}. The wide opening angle subtended by the
outflow limbs in the case of source I is indicative of a poorly
collimated driving source.  Thus our simulations begin with a
spherical wind driven into the grid via an inflow boundary condition.
This boundary condition is set by reestablishing wind conditions on a
``wind sphere'' in the grid before every time step. The wind impinges
on a collapsing, rotating molecular envelope. The density
distribution in the infalling ambient envelope are prescribed by
equations (8), (9) and (10) of the self-gravitating, rotating collapse
model of \cite{hartmann}. This infalling envelope is the same as that used in
previous outflow models of \cite{delamarter} and \cite{gardiner} where
it was shown that a combination of inertial and ram pressure
confinement was sufficient to collimate the wind into a bipolar
outflow.

The rotational velocity and infall speed of the envelope is chosen to
be appropriate for a $M = 10 M_{\sun}$ central gravitational source
\citep{ulrich}. Note that previous works \cite{delamarter,gardiner}
explored the interaction of infall and winds in the context of {\it
low mass} young stellar objects. The simulation parameters used here
are listed in table \ref{tparam}.  The wind consists of ionized H and
20\% atomic He by mass.  The ambient envelope is assumed to be
composed of $H_2$ and 20\% atomic He by mass. We note that even with
AMR methods the high cooling rates achieved behind the shocks on the
small scales at which these simulations are run ($\sim 50 ~AU$)
present a significant numerical challenge.  From the solutions to the
rotating collapsing sheet problem we find typical ambient densities of
order $n \sim 10^7 ~cm^{-3}$.  Because the cooling rate increases as
$n^2$, the cooling parameter in the flow, defined as $\chi =
t_{cool}/t_{hydro}$ is very small with $\chi << 1$ for most shock
conditions. Thus, in previous works many authors have chosen
isothermal equations of state such that $P \propto \rho^{\gamma}$ with
$\gamma \approx 1$.  Such a description can mimic certain aspects of
cooling, such as strong compressions behind shocks, but will not
correctly recover the dynamics in more complicated flow patterns. It
is more desirable therefore to explicitly track cooling when possible.

Note that in wind blown bubbles three discontinuities will form: a
``wind shock'' facing back into the freely expanding wind, an ambient
shock facing outward into the envelope and a contact discontinuity
between the two shocks delineating the interface between shocked wind
and shocked ambient material (figure \ref{schematic}). The strength of
cooling determines the distance between the shocks and the contact
discontinuity.  The principle difficulty in carrying forward
simulations such as those described here is achieving adequate
resolution to track the flow between the two shocks in the
wind-ambient material interaction region. The smallest scales which
can be captured with our runs is of the order $\Delta x \sim 0.2~AU$.
Thus cooling scale lengths must be of this order or not significantly
less than this if we are to capture the details of the post-shock flow
patterns.

In our simulations we have achieved a balance between realism and
numerical efficacy by modifying conditions in the wind and ambient
medium.  For example we have used an initial wind temperature of $T_w
\approx 10^4~K$ in the launch region which is likely too high.  We use
this value as it provides a mach number of $M \approx 20$ which is
useful for launching the simulation.  We allow the wind to cool as it
expands.  More importantly we have reduced the densities in the wind
and infalling cloud such that cooling plays a strong role but the
interaction regions can be resolved.  Thus while outflow rates of
$10^{-7}~M_{\sun}~yr^{-1}$ to $10^{-6}~M_{\sun}~yr^{-1}$ are
representative of higher mass YSO sources \citep{konigl} we have
chosen an outflow rate of $1.5 \times 10^{-9}~M_{\sun}~yr^{-1} $ to
allow our code to adequately resolve strongly cooled shear layers
present in the flow. If the wind were launched from an accretion disk
this would give a wind mass loss rate that was between $.1$ to $.01$
lower if standard disk wind theory can be applied. The discrepancy in
mass loss rates is clearly a significant difference between our models
and the actual situation in source I.  This difference should not
effect our principle conclusions as these are not sensitive to the
details of the cooling.

First we note that in our simulations the flow along the walls of the
cavity, at its base, {\it is} strongly cooling.  The principle
difference in this region of the flow between our simulations and one
with higher mass loss rates is the width of the wall and distance
between the wind shock and the ambient shock. We note that we are
interested primarily in the morphology of the outflow base of Source I
where the masers define the {\bf X} as well as the dynamics of
rotation in the swept-up material.  It has been shown that the
morphologies of wind blown bubbles are principally determined by the
ratio of specific momenta (or inertia) in the envelope to that in
the wind \citep{icke}. In our case, where we hold the stellar wind
velocity and stellar mass constant, it is the infall to outflow mass
loss rates ($f = \dot{M}_i/\dot{M}_w$) which determine the qualitative
details of the flow \citep{shu,delamarter}.  This is particularly
true along the arms of the flow at its base where higher mass loss
rates (and stronger cooling) will only narrow the opening angle by a
few degrees. We have observed this in our tests in which we have run
larger scale simulations with higher mass loss rates as well as cases
with an isothermal equation of state.  These have shown similar
results in terms of the shape of the outflow arms. It is the details
of smaller scale flow features, such as those associated with
instabilities in the swept up shell, which differ when the interaction
region is resolved.  With higher mass loss rates the walls of the
cavity become so thin that, even with AMR, they span only a few zones
and we are not able to resolve their internal dynamics.

\begin{table}[!h] \caption{Simulation Parameters. See text for details.
\label{tparam}}
 \begin{tabular}{l l}
 \tableline
 Wind Radius, $r_w$ & $20~AU$ \\
 Simulation Domain & $2.5~r_w$ radius by \\
 & $5~r_w$ along axis \\
 Computational cells per  $rw$ & 102 \\
 Wind Boundary Velocity, $v_w$ & $300~km~s^{-1}$ \\
 Wind Boundary Mass Flux, $\dot{M}_w$ & $1.5 \times 10^{-9}~M_{\sun}~yr^{-1}$ \\
 Infall Mass Flux, $\dot{M}_i$ & $1.5 \times 10^{-8}~M_{\sun}~yr^{-1}$ \\
 Wind Boundary Temperature & $1.8\times10^4~K$ \\
 Ambient Temperature & $50~K$ \\
 Central mass, $M_*$ & $10~M_{\sun}$ \\
 Collapse radius, $r_0$ & $100~r_w$\\
 Centrifugal radius, $R_c$ & $0.85~r_w$\\
 Flattening parameter, $\eta$ & $1.5$ \\
 Dynamical Age, $t_{dynam}$ & $7.1$ years \\
 \tableline
 \end{tabular}
\end{table}

\section{Results} The interaction of a spherical wind expanding into an
asymmetric density environment has been well studied both analytically
and numerically \citep{shu, icke, mellema, frankmellema}. For a review
see \cite{frank}.  In \cite{delamarter} the combination of inertial
confinement and ram pressure confinement from a collapsing envelope
was shown capable of producing a variety of bipolar outflow
configurations ranging from well collimated jet-like outflows to wider
butterfly shaped outflows with narrow waists. In \cite{delamarter} the
parameter $f = \dot{M}_i/\dot{M}_w$ was found to be critical in
determining the shape of the outflow where $\dot{M}_i$ and $\dot{M}_w$
are the infall and outflow rates respectively.  They found that $f =
10$ models produced fairly wide bubble with opening angle of
$60^\circ$. \cite{delamarter} also found that the post-shock cooling
produced second order effects on the flow.  In particular the
spherical wind will impinge upon inward facing ``wind shock'', (which
defines the inner walls of the outflow cavity), at an oblique
angle. This material will retain much of its initial velocity and will
be directed to stream along the contact discontinuity (CD) between the
shocked wind and shocked ambient material. Thus the CD becomes a
strong slip surface. Note also that the ambient material that passes
through the bow shock would be directed to flow toward the equator
while the shocked wind will flow toward the head of the outflow. There
has been some debate as to the nature of the mixing which occurs in
this region and which way the net flow of momentum will travel. Our
simulations provide some answer to these questions and these answers
are relevant to the nature of the Source I outflow.

\begin{figure}[!h]
\includegraphics[width=0.99\textwidth]{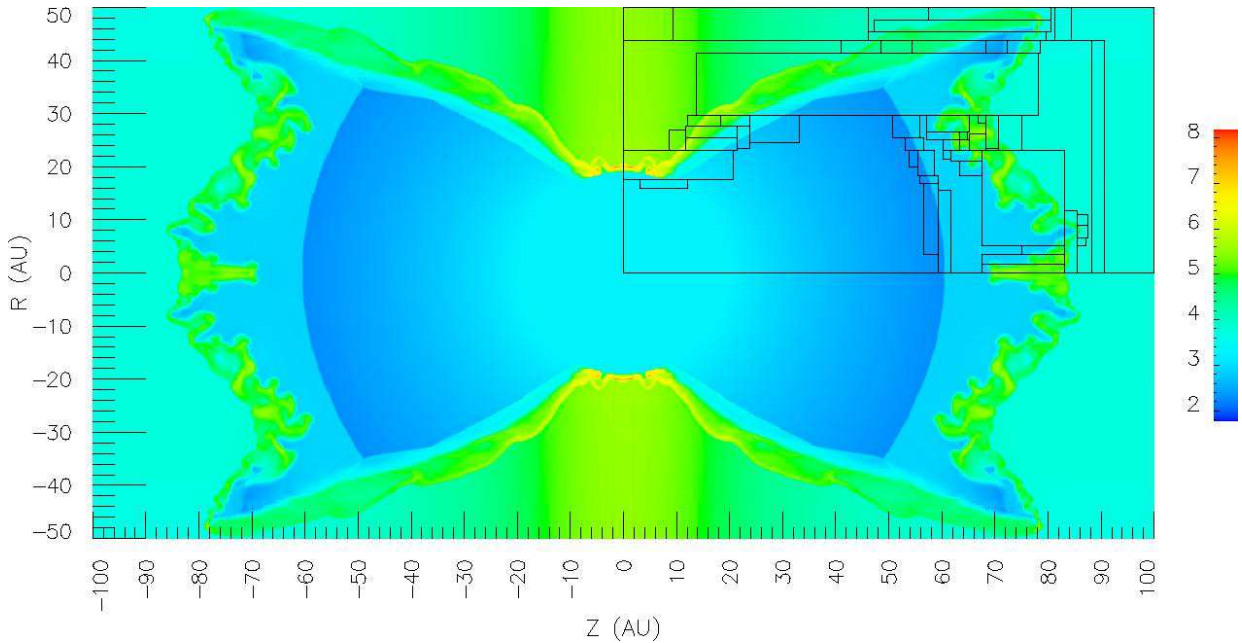}
\caption{Density plot in units of $log_{10}(n_{proton}~cm^{-3})$ at
time $t=7.1$ years.  The black ``hotboxes'' delineate regions of
enhanced refinement.
\label{den}}
\end{figure}

\begin{figure}[!h] 
\includegraphics[width=0.5\textwidth]{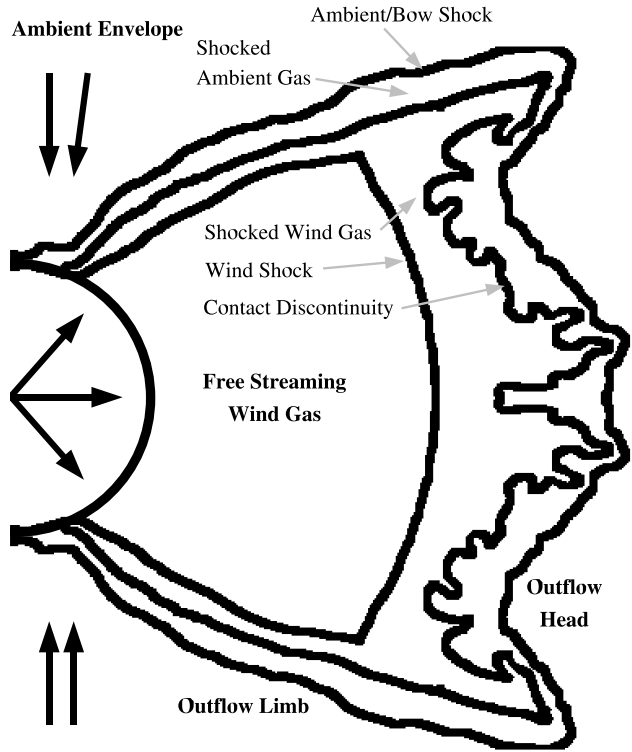}
\caption{Flow region and discontinuity schematic. \label{schematic}}
\end{figure}

Figure \ref{den} shows a density map of the outflow created in the
simulation. Globally we see the evacuation of a bipolar cavity where
the rotating ambient molecular material is swept toward the perimeter
of the cavity walls.  The global shape of the outflow is similar to
that seen in the studies of \cite{delamarter} who used an entirely
different code. This gives us confidence that the basic dynamics is
being correctly modeled.  The shell of swept-up material is thin due
to the strong energy losses from molecular dissociation and
cooling. The bulk of the cavity's volume is occupied by freely
expanding pre-shocked wind. Along the sides of the cavity we see that
once wind material strikes the inward facing wind shock its flow is
directed along the CD in a relatively thin shell. At higher latitudes
which represent the ``head'' of the outflow, the freely expanding wind
encounters the wind shock at a far less oblique angle and is more
strongly decelerated. The pressure retained in this region even after
cooling pushes the wind shock away from the CD (figure
\ref{schematic}).  We note that simulations with higher mass
infall/outflow rates but calculated on larger scales only show
differences at the head of the outflow as the post-wind shock material
is able to cool more effectively and moves closer to the CD.  We
briefly address the dynamics at the head of the outflow below.  For
now we note that the long term evolution of the outflow (in terms of
the cavity walls) relaxes to what appears to be a steady state as
inflow and outflow pressures balance.  Thus while the outflow head
dynamics is of general interest, the only part of the Source I outflow
which can be observed through the $SiO$ masers on $< 70~AU$ scales
will be the arms at the base of the cavity.

\begin{figure}[!h]
\includegraphics[width=0.99\textwidth]{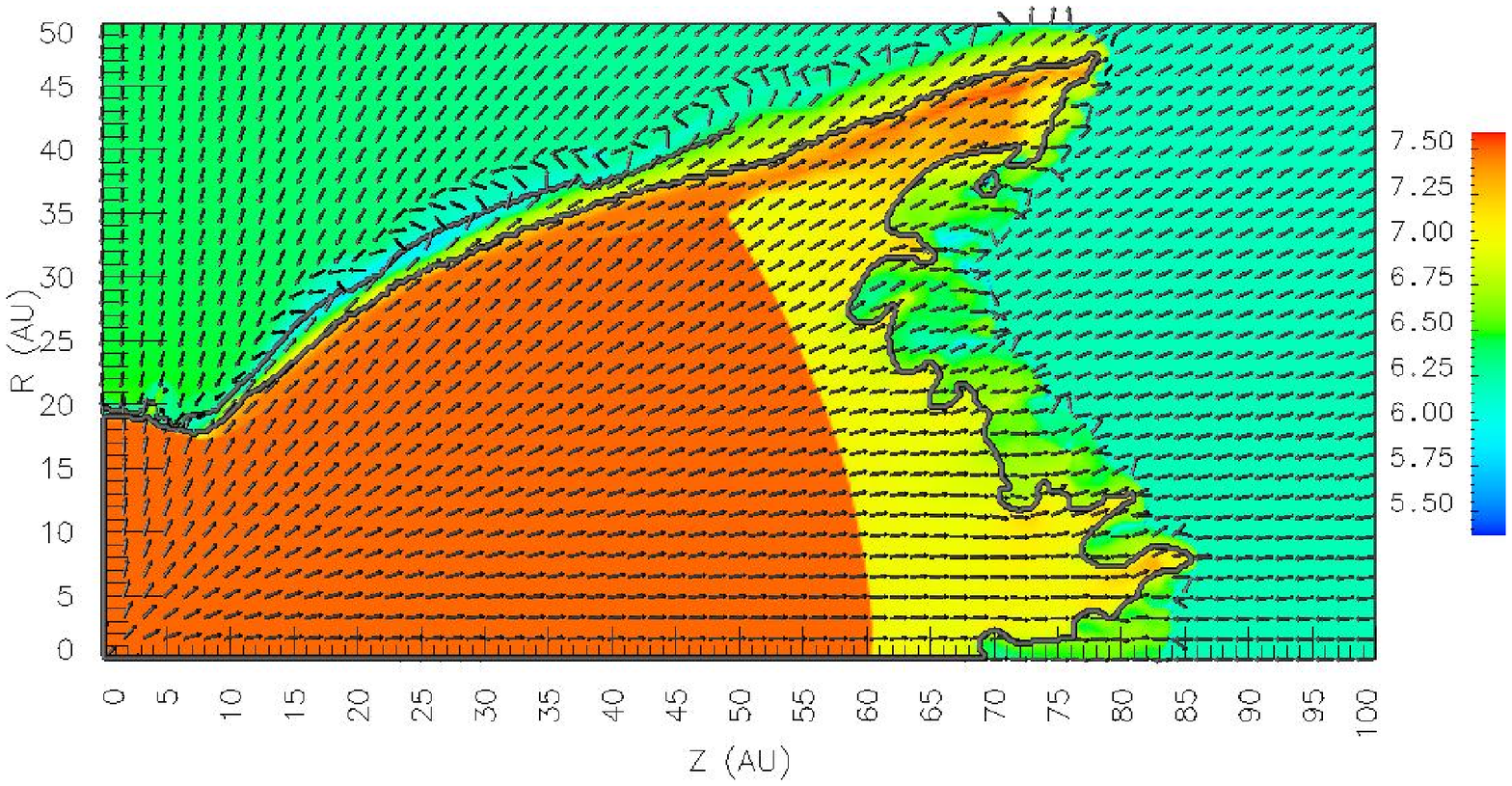}
\caption{Magnitude of the total velocity of the flow field in units of
$log10(cm~s^{-3})$ with arrows indicating flow direction.  A plot that
enhances the shear flow in the outflow limbs is shown in
figure\ref{vzoom}. The inner line delineates the contact
discontinuity.  The outer line delineates the location of the cross
cut in figure \ref{vrot}. \label{v}}
\end{figure}

\begin{figure}[!h]
\includegraphics[width=0.75\textwidth]{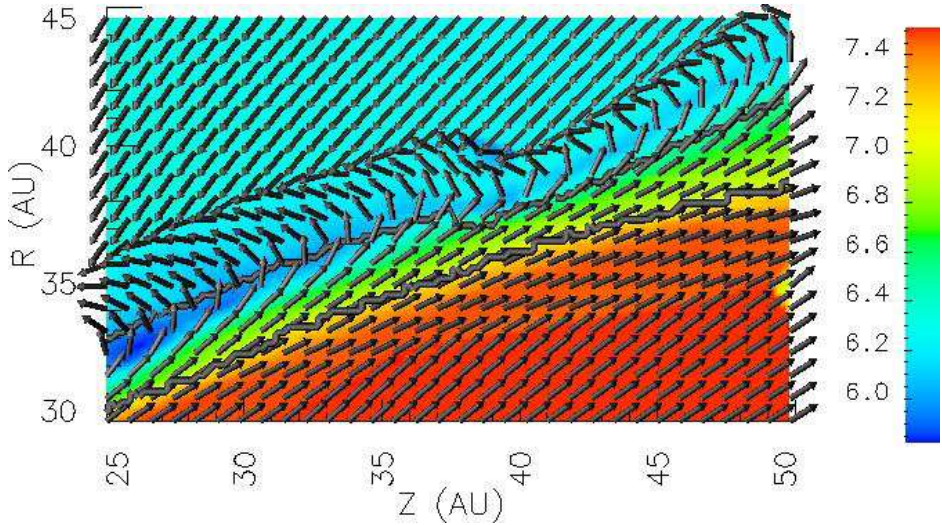}
\caption{An enhancement of the shear region in the outflow limb shown
in figure \ref{v}. \label{vzoom}}
\end{figure}

The most important aspect of these simulations for the subject at hand
is the fact that supersonic wind and ambient rotating molecular
material are compressed at the wind and ambient shocks forming a thin
layer around the contact discontinuity (figures \ref{den} \&
\ref{schematic}).  This layer then becomes a slip stream surface. The
morphology of the outflow is also crucial and is the result of the
ambient material's density and velocity distributions (i.e.  inertial
and ram pressure confinement). Figures \ref{v} \& \ref{vzoom} show the
kinematics of the flow with vectors indicating poloidal flow direction
overlaid on a map of velocity magnitude (we defer discussion of
rotation until the next section). Here one can see that wind material
strikes the inner shock at an oblique angle.  Post-shock wind is
focused into a flow that streams parallel to the contact
discontinuity.  This high speed shocked wind moves ahead of the rest
of the outflow, forming a cusp at mid to high latitudes that pushes
past of the head of the outflow at the poles. These cusps are
transient features and will eventually be subsumed into the rest of
the outflow. Dense limbs of wind swept ambient material delineate the
low latitude edges of the cavity. The dense limbs of the cavity and
the narrow waist formed close to the inflow boundary are essential
morphological signatures present in both our simulations and the
source I maser spot observations.

Velocity vectors of the flow pattern shown in figures \ref{v} and
\ref{vzoom} illuminate the formation of a slip stream between
initially infalling ambient molecular material and outflowing wind
material. The inner line in figure \ref{v} delineates the contact
discontinuity.  This marks the transition between mostly wind material
and mostly ambient material. Note that the change in direction of the
velocity vectors as one moves from ambient to wind material. This flow
reversal indicates the presence of a vortex across the interaction
region. We identify the large density and velocity gradients present
across the slip stream as susceptible to Kelvin-Helmholtz
instabilities and will discuss these in more detail in the next
section.

Thus to conclude this section we find our simulations show that a wind
blown bubble with appropriate morphology for Source I will form via
the interaction of a spherical wind with a collapsing, rotating
envelope.  In the next section we discuss both generic outflow
issues raised by the simulations as well as specific connections to
source I maser observations.

\section{Discussion} We break our discussion into two sections.  First we
review our results in light of previous simulations of wind driven
molecular outflows examining those features of the simulations which
shed new light on unresolved issues. In the second section we compare
our simulation results to the observations of Source I focusing
particularly on the rotational patterns.

\subsection{Generic Outflow Issues}

Several authors \citep{shu,wilkin,lee} have constructed analytic
models of the formation of bipolar molecular outflows.  These models
invoke the assumption that mixing between the shocked wind and shocked
ambient gas occurs instantaneously. Such rapid and local mixing yields
an outflow cavity that is delineated by a purely momentum driven thin
shell and the swept up wind mass is taken to be negligible. The last
assumption implies that the post-shock flow is dominated by shocked
ambient gas. Thus gas will be carried along the shell downward toward
the disk.  Our simulation contradicts this model assumption.  While
there is much more mass in the ambient material, the momentum in the
wind material is significant owing to the high velocity of the wind
and the oblique nature of the wind shock. We find that the momentum of
the wind material is dynamically important and must be incorporated
into any model used to predict the kinematics inside the thin
shell. This is true even in cases where turbulence dominates and the
cavity walls are fully mixed.  \cite{delamarter} derive a condition
for the mixed flow to be directed upward towards the head of the
outflow:
\[ \left[\frac{\dot{M}_w}{\dot{M}_i}
\left(\frac{\tilde{M}_{sw}}{\tilde{M}_{si}}\right)^2\right]>1 \]
where
$\tilde{M}_{sw}$ is the Mach number of the shocked wind material and
$\tilde{M}_{si}$ is the Mach number of the shocked infalling material.

\begin{figure}[!h]
\includegraphics[angle=-90,clip=true,width=0.75\textwidth]{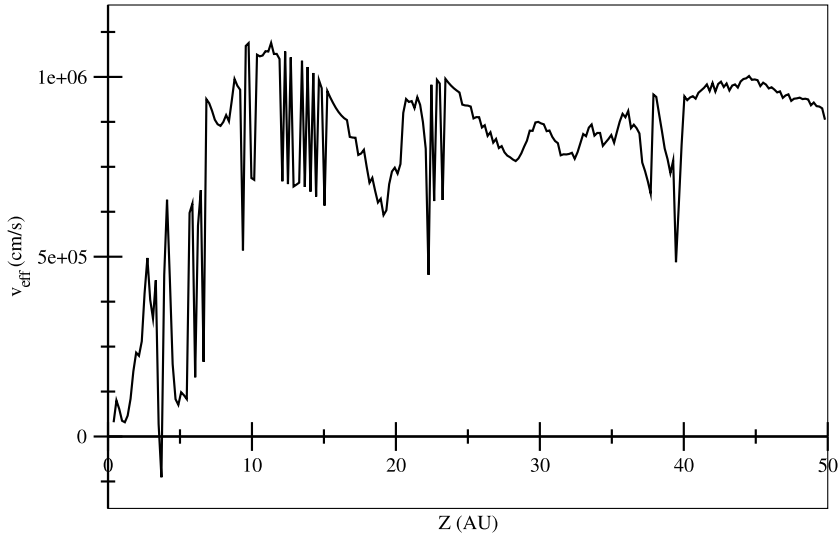}
\caption{Net velocity of fully mixed shocked material along the
direction of the contact discontinuity delineating the cavity
wall.\label{mmean}}
\end{figure}

For the parameters used in this simulation, $\dot{M}_i/\dot{M}_w=10$,
$\tilde{M}_{sw}=5$, $\tilde{M}_{si}=.1$. Thus the left hand side of
this equation evaluates to 250 and the condition for upward flow is
satisfied.  While the cavity walls in our simulations are not fully
mixed we can compute the direction the flow would take if mixing
occurred by averaging the momentum across the zones which comprise the
shell. Figure \ref{mmean} plots the effective velocity $v_{eff}$ that
would result along the direction of the outflow cavity wall if the
shocked layer were to become fully mixed.  Here positive velocity
indicates flow streaming upward towards the head of the cavity. Figure
\ref{mmean} shows the flow within the cavity walls to be positive at
essentially all latitudes with a velocity of $v \sim 8 ~km~s^{-1}$. This
result indicates that if the shocked flow were to mix by any means,
the momentum of wind material would overwhelm that of the slower,
denser ambient material resulting in poleward movement of material
along the shell.  Thus we conclude that the assumption that mixing
within the bipolar outflow shell leads to a net downward streaming of
mass towards the disk is incorrect.

Note that while do not see full mixing, our simulations do capture one
net turnover in flow direction.  We interpret the vortical reversal of
vectors between the ambient shock and the wind shock as due to the
development of Kelvin-Helmholtz instabilities and related mixing
processes. This mixing results in the entrainment of outwardly
directed wind momentum into the shocked ambient gas along the inner
cavity walls.  While our simulation clearly shows the vortical motions
in the interaction region we operate with limited resolution due to
computational constraints and cannot resolve the expected multiple
non-linear Kelvin-Helmholtz roll-ups in this region.  We rely,
therefore, on an analytic description to quantify the unresolved
instability responsible for entraining momentum from the wind into the
region of ambient gas. The characteristic growth rate of the
Kelvin-Helmholtz instability for wave mode $k$ is given by:
\[\Gamma_{kh} = |V_{sw}-V_{sa}|k\frac{\sqrt{\rho_{sw}
\rho_{sa}}}{\rho_{sw}+\rho_{sa}}\]
where $\rho_{sw}$ and $V_{sw}$ denote the density and velocity on the
shocked wind side of the slip stream and $\rho_{sa}$ and $V_{sa}$
denote the density and velocity on the shocked infalling side of the
slip stream. We have computed the time scale for the growth
Kelvin-Helmholtz modes $t_{kh}$ of wavelength $\lambda$, relative to
the dynamical time scale of the outflow source as $1/\Gamma_{kh} =
t_{kh} \approx 0.05 t_{dyn}\lambda/r_w$ where $r_w$ is the distance
from the central source to the wind shock. The analysis used here is
for the Kelvin-Helmholtz growth across a thin shear layer.  Since
wavelengths less than the thickness of the slip stream layer are
damped, we choose the wavelength of fastest growth equal to the width
of the cavity wall or shell $\lambda \approx 0.25 r_w$.  Using values
of the parameters taken from the simulations we find timescale for the
growth of this mode is much less than the dynamical timescale of the
outflow (i.e. $t_{kh} << t_{dyn} = 7.1~years$). This implies that
unresolved Kelvin-Helmholtz instabilities will provide the mixing
required to exchange momentum across the shear layer even though in
our simulations we see a smooth (numerically) diffusive mixing in the
shear layer.  This result supports our interpretation that mixing has
resulted in the entrainment of outwardly directed wind material into
the layer of shocked ambient gas.  With higher resolution we would
expect to see the shear layer become more turbulent.

We now address the fragmentation of the head of the outflow in our
simulations. While details of the polar caps of the outflow are not
illuminated by the Source I $Si0$ maser observations, the
fragmentation of the thin shell of swept-up ambient material in this
region warrants some discussion. The presence of these fragments
dominates the dynamics of the flow in the polar caps. We note that the
dynamical age of the outflow in our model is very young; 7.1 years.  A
typical astrophysical outflow of much greater dynamical age would have
time to produce even richer fragmentation and clumping spectrum than
our the flow modelled in our simulation. The instability along the
polar caps of the outflow and the formation of associated fragments
and inhomogeneities is interesting in that it is characteristic of YSO
molecular outflows \citep{mccaughrean}.  The nature of ``clumpy''
flows remains largely unexplored though \cite{polu, melioli} have
examined properties of inhomogeneities on the propagation of shocks
and the role such phenomena play in astrophysical contexts. Our
simulations imply that such clumpy flows are likely to be a generic
feature of bipolar outflows and jets.

We can, perhaps, understand the fragmentation seen in our simulations
through known unstable modes of shocks. \cite{vishniac} has examined
the stability of a thin spherical shock of wavelengths $>>$ than the
thickness of the thin shell. They find that shocks that are
sufficiently radiative to produce a density contrast of $\ga 10$ are
dynamically unstable.  The growth rate associated with these modes are
given by $\Gamma_{ts} \approx c_s/h$ where $c_s$ is the sound speed
inside the thin shell and $h$ is the shell thickness. The
characteristic temperature $10^4~K$ and shell thickness of $2~AU=0.1
r_w$ along the thin dense shell at the rapidly expanding polar caps in
the simulated outflow yield a timescale for the growth rate of this
instability of $1/\Gamma_{ts}= t_{ts} \approx 0.04 t_{dyn}$.  Thus the
rapid growth of the fragments which develop in this region are
consistent with Vishniac's analytic prediction.

\subsection{Comparison with Observations} In this section we make a
qualitative comparison between the broad morphological and kinematic
properties of our simulations and those observed in the $SiO$ maser
spots of Source I.  We note that more detailed quantitative
comparisons will require high resolution 3-D simulations including
radiation-transfer calculations beyond the scope of the current work.

First we note our previous works have also shown that the opening
angle of the outflow depends on the degree of flattening of the
ambient material (\cite{delamarter}. Flatter, more pancake-like
density (and infall velocity) distributions lead to more spherical
bubbles with a larger opening angle for the arms of {\bf X}.  Thus the
morphology of the models can be smoothly adjusted to fit the
conditions in Source I.  Most importantly the wide opening angle seen
in the Source I masers argues strongly for the presence of a
significant wide-angle wind component to the driving wind.  While a
jet component may still exist the laterial expansion of the lobes at
the base would be difficult to reproduce without a wind component with
significant momenta expanding into low latitudes.

\begin{figure}[!h]
\includegraphics[width=0.99\textwidth]{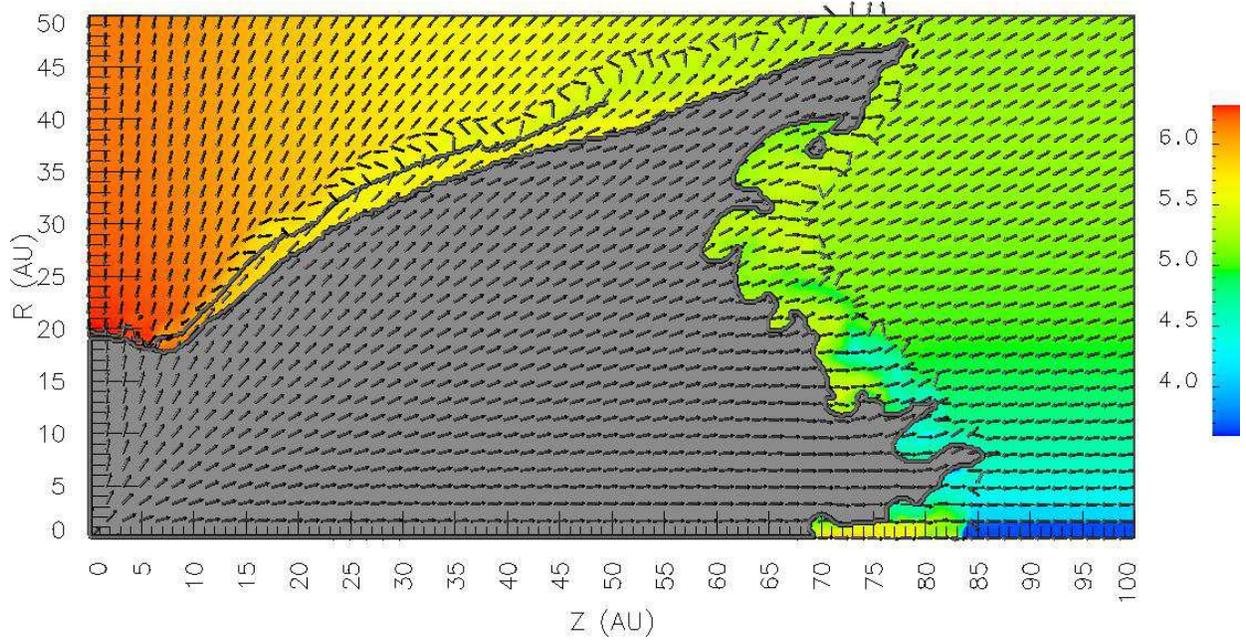}
\caption{Magnitude of particle rotational velocity $v_{\theta}$ of the
flow field in units of $log10(cm~s^{-1})$ with arrows indicating flow
direction.  The inner line delineates the contact discontinuity.  The
outer line delineates the location of the cross cut in figure
\ref{vrot}. \label{vrotmap}}
\end{figure}

\begin{figure}[!h]
\includegraphics[width=0.75\textwidth]{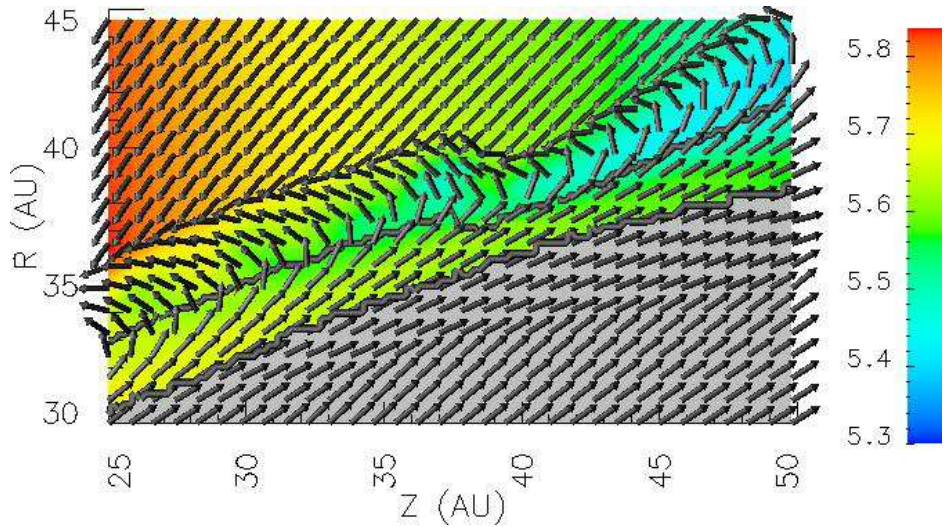}
\caption{An enhancement of the shear region in the outflow limb shown
in figure \ref{vrotmap}. \label{vrotzoom}}
\end{figure}

We have proposed that the line of sight maser velocities observed
about Source I are due to rotation.  Specifically it is rotation
retained by infalling molecular material that has been intercepted by
a biconical outflow.  We now demonstrate this assertion. Figures
\ref{vrotmap} \& \ref{vrotzoom} show a color map of the rotational
component of velocity $v_{\theta}$ in the shell of swept up
material. The non-rotating wind material is delineated by the interior
line. Notice that the region of greatest shear occurs midway between
the outer edge of the ambient shock and the slip surface (CD). This is
where the flow undergoes direction reversal across the slip stream. As
discussed above we can identify this region as adjacent to the line of
greatest vorticity generation.  This region is most susceptible
Kelvin-Helmholtz instabilities and the creation of related density
inhomogeneities. The formation of fragments and filaments in this
region would provide the density and path length enhancement most
likely to result in the amplification of master spots. Because the
masing material most likely lies in this region, its kinematics would
be likely be revealed by observations of maser spots.

\begin{figure}[!h]
\includegraphics[angle=-90,clip=true,width=0.75\textwidth]{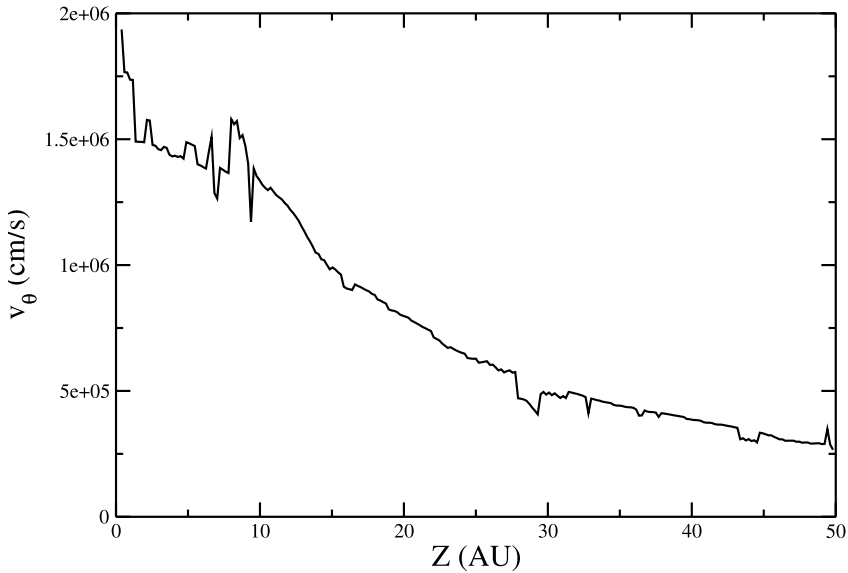}
\caption{Rotational velocity vs distance along the z-axis taken along
the surface midway between the position of highest density in the
outer shell of shocked ambient material and the contact surface
separating the wind launched and shocked ambient gases along the
r-axis.  The position of this cut is show in figures \ref{v},
\ref{vzoom} \ref{vrotmap} \& \ref{vrotzoom}. \label{vrot}}
\end{figure}

Figure \ref{vrot} plots a cut of rotational velocity taken along this
surface midway between the ambient shock and the contact surface. The
magnitude of the rotational velocities is $3~km~s^{-1} < v_{rot} <
15~km~s^{-1}$ in rough agreement with the line of sight velocities
seen in the maser data.  These values reflect the rotational
velocities of the pre-shock ambient material. Thus as ambient material
spirals inward towards the star it intercepts the cavity wall, is
shocked and eventually reverses its poloidal but not its toroidal
velocity. Our results thus indicate that the observed line of sight
velocity of the maser spots can be interpreted as the rotation of
infalling material that has been intercepted by a poorly collimated
biconical outflow. Note that we correct for a systematic $\sim 5~km~s^{-1}$
redshift as is appropriate for the Orion region.

Note also that our results show a decrease in rotational velocity as
one moves outward along the arms of the {\bf X} defined by the maser
spots.  The presence of a velocity gradient is ambigious in the
observations. For example, in the data of \cite{greenhill98}, the
redshifted lobes show a velocity gradient with values ranging from $25
~km~s^{-1}$ at the base of the {\bf X} to $15 ~km~s^{-1}$ at its
farthest extent. The blueshifted lobes do not show such a clear
gradient however they do show a similar range of velocities.  In our
model gradients in $v_{rot}$ in the maser spots will reflect, to first
order, gradients in the rotational velocities of the ambient material.
A better accounting of turbulent advection of material along the walls
of the cavity could smooth out the observed rotational
gradient. However since no torques act on the swept up material
conservation of angular momentum will still reduce its rotational
velocity.  Consideration of the initial conditions shows that the
gradient in $v_{rot}$ is stronger in the radial direction than in
height. Thus for models with wider opening angles the we expect the
gradient to be less dramatic.

Finally one may ask if the rotation inferred for the Source I masers
is due to the ambient material as we have modelled or if it comes from
rotation of the stellar wind.  We note that the observed maser proper
motions of $\sim 15~km~s^{-1}$, which we take to be poloidal
velocities along the shell, are an order of magnitude smaller than
both typical young stellar object outflows and stellar winds.  If this
material were to originate from a disk wind it would have to emerge
from a region quite far from the central source.  Using $v_{escape}
\sim v_{Keplerian}(R)$ gives $R \sim 40~AU$ which is quite far from the
inner regions of the disk where launching is expected to be most
effective \citep{ouyed}. One might argue that the modest rotational
gradient observed in the maser data indicates the presence of a
magnetic field in the wind anchored to the disk such as would be the
case for a disk wind.  The field would then sustain the rotational
motions of the wind material caught-up in the masers.  However once
any MHD launching mechanism takes a wind parcel out beyond the
Alfv\'en radius the field will no longer provide rotational support
for the wind and angular momentum conservation will reduce the wind's
rotational velocity as it expands.  Since the Alfv\'en radius tends to
be a few times larger then the radius of the footpoint of the flow,
and the flow is likely to form close to the star at the inner regions
of the disk, it is unlikely that the winds rotational motion can be
magnetically supported. Thus the rotation seen in the masers are most
convincingly interpreted as ambient material that has been swept up or
entrained along shocks with the fast-moving wind \citep{greenhill,
doeleman}. In all models, the $H_2$ densities and temperatures
necessary for maser action, and the small line-of-sight velocity
shifts necessary for maser amplification occurs along the limbs of the
outflow.

\section{Conclusion}
Using AMR simulations we have explored the evolution of an outflow
driven by a spherical wind from a massive gravitating source
interacting with a rotating infalling cavity.  The dynamics of the
flow in the shocked ambient material are essential aspects of our
model that have been neglected in previous works. Specifically, the
growth of Kelvin-Helmholtz unstable modes in the slip stream between
outflowing wind material and infalling molecular material along the
cavity walls provides mixing at the wind/ambient gas interface.  This
mixing results in the outward acceleration and eventual direction
reversal of initially infalling ambient material.

We have also shown that the head of the outflow will be unstable to
thin shell instabilities.  While this conclusion is not relevant to
the observations of Source I it is of general interest for studies of
molecular outflows. The fragmentation of radiatively cooling outflow
lobes has important consequences for their long term dynamics and
observational properties.  The fact that our simulations show a rapid
transition to fragmentation implies that molecular outflows on larger
scales can be expected to be ``clumpy'' on a variety of scales. This
issue should be addressed in future works.

We have compared our simulations with observations of maser spots in
Source I in the BN/KL region in Orion.  We find that when the wind
evacuates a bipolar cavity, swept-up rotating ambient molecular
material in the cavity walls is the likely source of the masers. This
zone contains material which retains its rotation but which has been
accelerated upward towards the head of the outflow lobe.  The poloidal
velocities seen in our simulations are of order those seen proper
motions in the observations.  The rotation retained by the shocked
ambient material is consistent with the observed line of sight
velocity of the maser spots observed about Source I. Thus we conclude
that the line of sight motions in Source I inferred to be due to
rotation \citep{greenhill, greenhill05} can be interpreted as
originating in the rotation of the collapsing ambient material. Future
work will need to explore the flow pattern in 3-D as well address
issues related to formation of the masers in greater detail.  We also
conclude that the wide opening angle of the maser spot pattern is
strong evidence that a significant wide angle wind component is at
work in the Source I outflow.
\acknowledgments We thank Lincoln Greenhill and Mark Reid for fruitful
conversations which helped this work.  Support for this work was
provided by NFS grant AST 00-98442, NASA grant NAG5-8428, an HST
grant, DOE grant DE-FG02-00ER54600, and the Laboratory for Laser
Energetics.

\appendix
\section{$H_2$ Microphysics}

\subsection{Cooling Functions}
\begin{figure}[!h]
\includegraphics[width=0.75\textwidth]{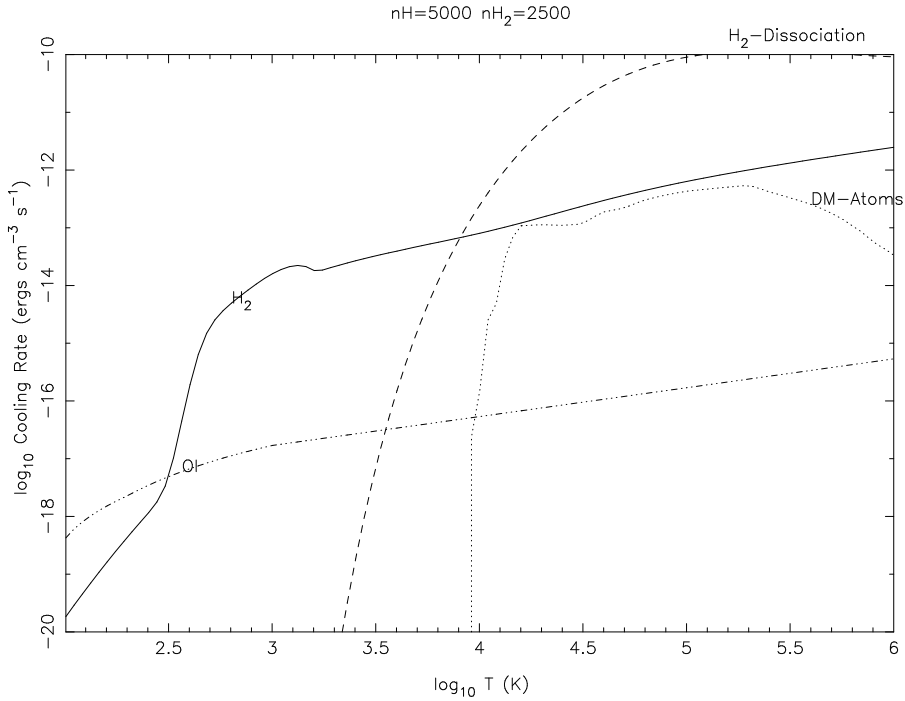}
\caption{Cooling rates for typical ISM abundances,
$n_{H_{2}}=2500$,$n_{HI}=5000$. $H_2$ is the molecular hydrogen
cooling function, $H2-Dissociation$ is the thermal energy loss due to
the dissociation of $H_2$ molecules, $OI$ is the singly ionized oxygen
line cooling and $DM-atoms$ is the atomic line and recombination
cooling function of Dalgarno \& McCray.
\label{cooling}}
\end{figure}

We employ a total cooling function $\Lambda = \Lambda_{LS} +
\Lambda_{DM} + \Lambda_{OI} + \Lambda_{HeI_D} + \Lambda_{HI_D} +
\Lambda_{H_{2\,D}}$.  Where $\Lambda_{DM}=n_{H_{ion}}n_{e^-}\lambda_{DM}$
and $\lambda_{DM}$ is the atomic line cooling function appropriate for
interstellar gases by \cite{dm}, the dominant cooling process at $T >
10^4$. $\Lambda_{LS}$ is the $H_2$ cooling function of \cite{ls}
appropriate for a molecular gas.  $\Lambda_{OI} = n_{HI} n_{OI}
\lambda_{OI}(T)$ dominates the cooling at $T < 300~K$ where
$\lambda_{OI}(T)$ is the OI line cooling function tabulated by
\cite{lr}. For temperatures greater than those listed in the table,
$\lambda_{OI}(T)$ extrapolates with $\sqrt{T}$. Because the first
ionization potential of $O$ is comparable to that of $H$, we use the abundance of molecular and atomic $H$ as a tracer for OI
given as $n_{OI} = (n_{H_2}+n{HI})/n_{H_{ions}} f_{OI}$ where $f_{OI} = 8.51
\times 10^{-4}$ is the fractional coronal abundance of oxygen.  The
$\Lambda_{HeI_D}$, $\Lambda_{HI_D}$, and $\Lambda_{H_{2\,D}}$ terms
account for cooling due to dissociation and ionization.
\begin{eqnarray*}
\Lambda_{HeI_D}   & = & 24.60 eV D_{HeI} \\
\Lambda_{HI_D}    & = & 13.59 eV D_{HI} \\
\Lambda_{H_{2\,D}} & = & 4.48 eV D_{H2} \\
\end{eqnarray*}
Where D represents the dissociation rates given in the following
sections.  Figure \ref{cooling} shows each of
these cooling rates for typical ISM density.

\subsection{Dissociation}
\cite{ls} fit an analytic function for the $H_2$ dissociation rate
which takes the form:
\[ \log_{10}(k^{H_2,H}_D) = \frac{\log_{10} k_H - \log_{10}(k_H/k_L)}{1+n/n_{cr}} \]
where, for $H-H_2$ collisions,
$ \log_{10}(n_{cr}) = 4.00 - 0.416x - 0.327x^2 $
and, for $H_2-H_2$ collisions,
$ \log_{10}(n_{cr}) = 4.13 - 0.968x + 0.119x^2 $
with $x=\log_{10}(T/10^4 K)$. $k_H$ and $k_L$ refer to the high ($n >>
n_{cr}$) and low ($n << n_{cr}$) density limits for the reaction rate.
\cite{ls} give:
\[ k_H(T)= \left\{ \begin{array}{ll}
    3.52\times10^{-9}\exp(-4.39\times10^4/T)\quad\textrm{for}\quad H-H_2 \\
    5.48\times10^{-9}\exp(-5.30\times10^4/T)\quad\textrm{for}\quad H_2-H_2 \\
\end{array} \right. \]
\cite{lrrw} Improve the dissociation low density dissociation rate of
\cite{ls} by using the more recent calculations for $k_L^{H-H_2}$ of
\cite{dove}. \cite{lrrw} fit the results of \cite{dove} to the form:
\[ k_L(T)= 4.69\times10^{-14}T^{0.746}\exp(-5.55065\times10^4/T) \quad\textrm{for}\quad H-H_2 \]
We modify the original rates further by using the more recent
calculations for low density $H_2-H_2$ dissociation of \cite{mk}.  We
have also included corrections to the dissociation rate through the
action of $He-H_2$ and $e^--H_2$ collisions of \cite{mk} given by:
\[ k^i_D(T)=\left( \frac{8dkT}{\pi\mu} \right) ^{1/2}
  \frac{a(kT)^{b-1}\Gamma(b+1) \exp(-Eo/kT)}{\left( 1+CkT \right)^{b+1}}
  \quad\textrm{for}\quad H_2-H_2 \]
where $d=1.894\times10^{-22}$ $k=3.167\times10^{-6}$ and the constants
a, b, c and Eo for each collision partner are given in table\ref{tmk} and
$\mu$ is the reduced mass of the collision pair.
\begin{table}[!h]
\caption{Dissociation Rate Constants. \label{tmk}}
  \begin{tabular}{lrrrr}
    \tableline
    Partner & a & b & c & Eo \\
    \tableline
    $H_{2}$ & 40.1008 & 4.6881 & 2.1347 & 0.1731\\
    He & 4.8152 & 1.8202 & -0.9459 & 0.4146\\
    $e^{-}$ & 11.2474 & 1.0948 & 2.1382 & 0.3237\\
    \tableline
  \end{tabular}
\end{table}
The total dissociation rate is given by:
\[ D_{H_2} = \sum_{i} n_i n_{H_2} k^i_D \]
where i ranges over all collision partners: $H_2$,$H$,$He$,and $e^-$.

\subsection{Molecular Recombination Rate}
\cite{hm} give an approximation to the $H_2$ recombination rate due to
HI ``sticking'' on dust surfaces as:
\begin{eqnarray*}
R_{H_2} & = & n_{H_{nuclei}}\,n_{HI}\, 3 \times 10^{- 17} cm^3 s^{-1} \frac{T_{2}^{1/2} f_{a}}{1 + 0.4(T_{2}+T_{dust\,2})^{1/2}+0.2T_{2}+0.08T_{2}^{2}} \\
T_{2} & = & T/100 \\
f_{a} & = & 0.5 = \textrm{ fraction of molecules that do not evaporate on dust surfaces.} \\
\end{eqnarray*}

\subsection{Shock Dissociation}
\begin{figure}[!h]
\includegraphics[width=0.75\textwidth]{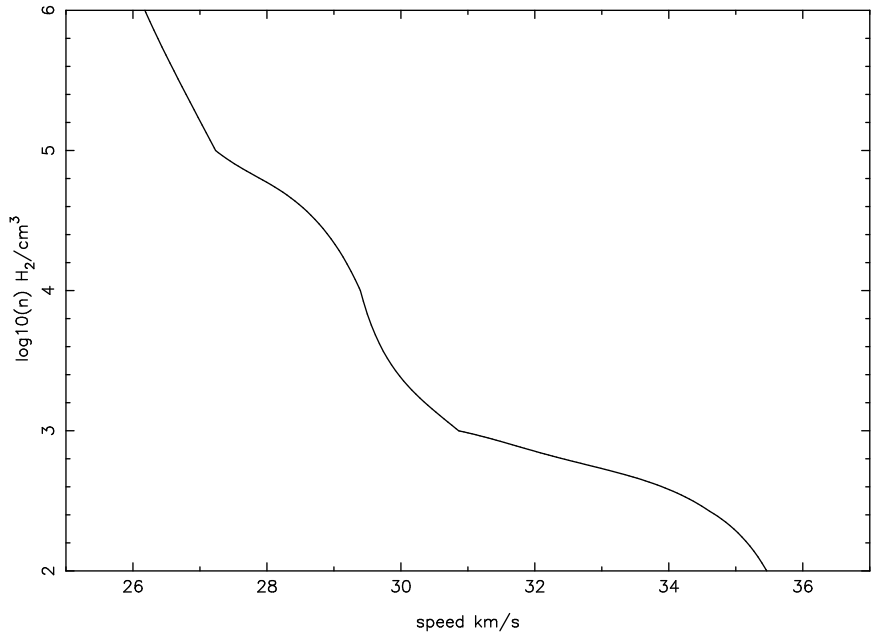}
\caption{Pres-hock density vs. shock speed for a steady shock resulting
in 90\% downstream $H_2$ dissociation. \label{sspeed}}
\end{figure}

As a test of these dissociation rates as implemented in our code we
have computed the impulsively launched steady shock speed required for
molecular dissociation.  Figure \ref{sspeed} shows pre-shock density
vs. shock speed for a steady shock resulting in 90\% downstream $H_2$
dissociation.  This result is consistent within a few $km~s^{-1}$ with
the results of previous authors \citep{smith,hm2}.

\subsection{Ionization \& Recombination}
\cite{m} catalog ionization and recombination rates for many species.
We have employed HeI and HI ionization rates originally from
\cite{ar}.  \cite{vf} fit the radiative recombination rate
coefficients for several species including the He and H recombinations
rates used for this work.  We have included the fit to dielectric
contribution to the He recombination rate of \cite{m}.

\end{document}